\documentclass[prd,tightenlines,nofootinbib,superscriptaddress]{revtex4}

\usepackage{amsfonts, amssymb, amsthm, amsmath, bbm, float, psfrag, graphicx}

\newcommand{\C}{{\mathbb C}}
\newcommand{\N}{{\mathbb N}}
\newcommand{\R}{{\mathbb R}}

\newcommand{\cD}{{\mathcal D}}

\newcommand{\cG}{{\mathcal G}}
\newcommand{\cH}{{\mathcal H}}

\newcommand{\cK}{{\mathcal K}}

\newcommand{\cP}{{\mathcal P}}
\newcommand{\cS}{{\mathcal S}}

\newcommand{\cV}{{\mathcal V}}
\newcommand{\cZ}{{\mathcal Z}}

\newcommand{\SU}{\mathrm{SU}}

\newcommand{\Spin}{\mathrm{Spin}}

\newcommand{\dsty}{\displaystyle}

\newcommand{\id}{\mathbb{I}}

\newcommand{\be}{\begin{equation}}
\newcommand{\ee}{\end{equation}}
\newcommand{\ba}{\begin{array}}
\newcommand{\ea}{\end{array}}
\newcommand{\bes}{\begin{eqnarray}}
\newcommand{\ees}{\end{eqnarray}}

\newcommand{\la}{\langle}
\newcommand{\ra}{\rangle}

\newcommand{\f}{\frac}

\def\nn{\nonumber}

\def\hH{\hat{H}}
\def\hcH{\hat{\cH}}
\def\tg{\widetilde{g}}
\def\tsigma{\widetilde{\sigma}}

\def\dsty{\displaystyle}

\begin{document}

\title{Effective Hamiltonian Constraint from Group Field Theory}

\author{\bf Etera R. Livine}\email{etera.livine@ens-lyon.fr}
\affiliation{Laboratoire de Physique, ENS Lyon, CNRS-UMR 5672, 46 All\'ee d'Italie, Lyon 69007, France}
\author{\bf Daniele Oriti}\email{doriti@aei.mpg.de}
\affiliation{MPI f\"ur Gravitationsphysik, Albert Einstein Institute,  Am M\"uhlenberg 1, D-14476 Potsdam, Germany}
\author{\bf James P. Ryan}\email{james.ryan@aei.mpg.de}
\affiliation{MPI f\"ur Gravitationsphysik, Albert Einstein Institute,  Am M\"uhlenberg 1, D-14476 Potsdam, Germany}

\date{\today}

\begin{abstract}

Spinfoam models provide a covariant formulation of the dynamics of loop quantum gravity. They are non-perturbatively defined in the group field theory (GFT) framework: the GFT partition function defines the sum of spinfoam transition amplitudes over all possible (discretized) geometries and topologies. The issue remains, however,  of explicitly relating the specific form of the group field theory action and the canonical Hamiltonian constraint. Here, we suggest an avenue for addressing this issue. Our strategy is to expand group field theories around non-trivial classical solutions and to interpret the induced quadratic kinematical term as defining a Hamiltonian constraint on the group field and thus on spin network wave functions. We apply our procedure to Boulatov group field theory for 3d Riemannian gravity. Finally, we discuss the relevance of understanding the spectrum of this Hamiltonian operator for the renormalization of group field theories.

\end{abstract}

\maketitle


\section*{Introduction}

The {\it Spinfoam} program \cite{alex} has been originally developed in order to implement, in a covariant, sum-over-histories form, the dynamics of loop quantum gravity \cite{thomas, carlo} and compute transition amplitudes between its spin network states of quantum geometry. However, the spinfoam formalism has been later found to be much more general, being intimately tied to the quantization of topological field theory (of the BF type) and very natural even in a lattice gauge theory context. The basic setting is that the space-time structure is described by an abstract 2-complex dressed with algebraic data given by representations and intertwiners (invariant tensors) of the gauge group (usually $\SU(2)$ or the Lorentz group $\Spin(3,1)$ for quantum gravity). Then a spinfoam model defines a probability amplitude for each such discrete space-time structure. In most models, the 2-complexes are (topologically) dual to space-time (pseudo-)triangulations and spinfoams can be interpreted as quantized simplicial geometries. From this point of view, spinfoam amplitudes are closely related to discretized general relativity and Regge calculus \cite{alex}; the relation between spin foam amplitudes and simplicial gravity path integrals has been known from early on \cite{alex}, and has been recently clarified further \cite{eteravalentin, valentin, gft_aristide,diffeos}. Finally, the full spinfoam quantum dynamics is defined as the sum over all possible 2-complexes and a non-perturbative definition of this sum is provided by the {\it group field theory} formulation \cite{gft0,gft1,gft_dan}.

A group field theory is a field theory on a group manifold (the product of a number of copies of the relevant gauge group) and with a peculiar type of non-local interaction. Its main characteristics is that its Feynman diagrams can be mapped onto 2-complexes and the associated Feynman amplitude of these Feynman diagrams define the spinfoam amplitude of the corresponding 2-complex. Then the perturbative expansion of the partition function of the group field theory (GFT) defines the sum of the spinfoam amplitudes over all (admissible) 2-complexes. From this perspective, group field theory can be considered as a generalized matrix/tensor model \cite{mm,tensor}, who generates (pseudo-)triangulations of space-time as Feynman diagrams (see the cited literature for further details).
Nevertheless, the particularity of group field theory is that it is a field theory in its own right and we can use standard field theory techniques to investigate and analyze its properties and its quantization (in particular, tools from QFT perturbative renormalization\cite{gft_renorm,gft_recent}. In fact, whatever their historic origin, one may take group field theory as an independent arena for research in its own right and in fact it is in this vein that a large portion of research on the subject is done nowadays \cite{gft_recent}.  Of course, the original motivations still stand, and, with this in mind, we shall attempt here to refocus on one of the initial goals:  to provide a consistent quantum dynamics for canonical loop quantum gravity states.

\medskip

Indeed, we propose to interpret the kinetic term of the group field theory, in the effective dynamics around a non-trivial back ground solution, as a Hamiltonian constraint acting on the group field and more generally on spin network states. 
The rationale for this proposal lies in the fact that group field theories  can be interpreted as second quantizations of spin network dynamics, thus as a sort of \lq 3rd quantization\rq of gravity \cite{gft1,gft_dan,gft_3rd}. As such one expects the dynamics of the 1st quantized theory, here canonical loop quantum gravity, to be encoded in the classical action of the 2nd quantized theory, here group field theory. Recall also that this dynamics can be defined either in terms of graph changing (from the point of view of spin network states) or in terms of non-graph-changing Hamiltonian operators \cite{thomas}. Moreover, in such a framework, one expects non only the geometry of spacetime to be fully dynamical, but also its topology. Group field theories fulfill these expectations and indeed incorporate, in their perturbative expansion, both a sum over all 2-complexes/triangulations of given topology, and a sum over all topologies. However, they do so in a peculiar way, as we are going to discuss: because of the trivial kinetic term usually chosen, the whole quantum dynamics of both geometry (Hamiltonian constraint) and topology is encoded in the interaction term of the theory, whose repeated action on spin network states (when seen as an operator) generates both a graph-changing evolution of geometry and a change in the underlying topology of space. What we would like to have, instead, is a non-trivial kinetic term that could be held entirely responsible for the dynamics of geometry, leaving the change of topology confined to the GFT interaction, possibly alongside additional contributions to the dynamics of geometry (see \cite{gft_3rd} for a discussion of this issue). We achieve this for the effective GFT dynamics. Moreover, we will see that the effective Hamiltonian constraint we generate will be of a non-graph-changing type.

Let us also mention that the exploration of the non-perturbative sector of GFT models, and of their effective dynamics around background configurations,  has already proceeded along different directions, recently \cite{gft_winston, noi, matrix, daniele_emergent, emergentmatter, danielelorenzo}.  In some works \cite{gft_winston,noi, matrix, daniele_emergent, emergent matter} the idea being investigated was that some simplified GFT perturbations around classical solutions could be interpreted as emergent (non-commutative) matter fields. Another possibility being explored  \cite{danielelorenzo} was to obtain effective equations for geometry from the GFT equations of motion, as conditions for a given background configuration to define a solution of the same, in the spirit of mean field theory in Bose condensates. Here we explore the other logical possibility that the dynamics of geometry should be looked for in the effective dynamics for generic perturbations around background solutions.

\

There are two main ingredients to our proposal.

First, we focus on the free group field theory defined by the quadratic part of the GFT action. The equation of motion of this free GFT is a linear equation of the type $\hH\,\phi=0$ where $\phi$ is the field and $\hH$ can be interpreted as a Hamiltonian constraint. However, the standard formulation of GFTs uses a trivial kinetic term and is of the type:
\be
S_{GFT}[\phi]=\f12\int \phi^2 -\lambda\int \cV[\phi],
\ee
where $\cV[\phi]$ defines the interaction term. Obviously, the free GFT defined as such is trivial. Our strategy here is to follow the procedure first used in \cite{gft_winston}. We can expand the GFT around a non-trivial classical solution $\phi_0$ to its full equation of motion:
\be
\phi_0=\lambda\f{\delta \cV}{\delta\phi}[\phi_0]\,.
\ee
This classical solution defines a background structure for the GFT\footnote{Notice also that generically these solutions are purely non-perturbative configurations as it is testified by their dependence on the GFT coupling constant.} and we can define an (effective) action describing the field variations around $\phi_0$ (instead of describing its variations around the ``no-space" state $\phi=0$):
\be
S_{\phi_0}[\phi]\,\equiv\,S_{GFT}[\phi_0+\phi]-S_{GFT}[\phi_0]
=
\f12\int \phi \,\hH_{\phi_0}\phi +\dots
\ee
The kinetic term is then non-trivial and provides us with a tentative Hamiltonian constraint for our proposal. This operator $\hH_{\phi_0}$ depends on the background structure defined by the field $\phi_0$, which encodes the full some non-trivial dynamical information since its definition involves the fundamental interaction term $\cV$.

The second ingredient is to view spin network functionals $\psi$ as multi-particle states of the GFT. They are indeed be constructed as the group-averaged tensor product of group fields, as we will explain in more detail:
$$
\psi\sim\phi\otimes..\otimes\phi\,.
$$
From this point of view, the Hilbert space of spin networks can be seen as a Fock space of the quantized GFT.
Then the linear operator $\hH_{\phi_0}$ acts on states $\psi$ and we can investigate its spectrum  on spin network states.

Finally, we have underlined the interpretation of the free GFT and the role of the kinetic term $\hH_{\phi_0}$ as defining a constraint operator acting on the group field $\phi$ and spin network functionals $\psi$. However, more generally, $\hH_{\phi_0}$ defines the (inverse of the) propagator for the point of view of quantum field theory and it is crucial to understand its properties and spectrum for the computation of the GFT correlations (which define the spinfoam transition amplitudes) and the renormalization of the GFT (which reflects the coarse-graining of spinfoam models). In fact, as natural in field theory context, and in 3rd quantized gravity \cite{gft_3rd}, the presence of interactions will necessarily involve excitations of quantum geometry outside the space of solutions of the Hamiltonian constraint (\lq virtual\rq, \lq off-shell\rq  geometries akin to virtual particles in ordinary field theory) and their understanding requires then a control over the full spectrum of the Hamiltonian constraint.

\medskip

The present paper consists in two parts. A first section will review the basics of the group field theory formalism. We will introduce its expansion around non-trivial classical solutions and discuss how to define spin network functionals as multi-particle states. In the second section, we will explicitly apply our program to Boulatov's GFT for the Ponzano-Regge spinfoam model of 3d quantum gravity. We will expand it around the flat solutions introduced in \cite{gft_winston} and analyze the spectrum of the induced Hamiltonian constraint. We will compare it to known Hamiltonian constraint of topological BF theory and we will see that it can be interpreted as a Klein-Gordon-like operator with a spectrum of the type ``$p^2+m^2$". We will finally conclude discussing the relevance of our procedure to the study of group field theories and their interpretation as quantum gravity models.

\section{Group Field Theory for SpinFoams}

\subsection{Generating the Spinfoam Partition Function}

A group field theory (GFT) is defined by the choice of a gauge group $\cG$ and an action  of the form:
\be\label{rev01}
\cS_\lambda[\phi]=
\f12\int[dg_a][d\bar{g}_{a}]\; \phi^\sigma(g_a)\;\cK(g_a, \tg_a)\; \phi^{{\tsigma}}(\tg_a)
-\lambda\int \Big(\prod_{a=1}^m [dg_{ab}]\; \phi^{\sigma_a}(g_{ab})\Big)\,\cV(\{g_{ab}\}).
\ee
where $\phi$ is a real (or complex)-valued function acting on $n$ copies of the group manifold $\cG$:
\be\label{rev02}
\phi: \cG^{\otimes n} \rightarrow \R \;(\textrm{or}\; \C) ; \quad (g_1, \dots,g_n) \rightarrow \phi(g_1, \dots, g_n) =: \phi(g_a).
\ee
The label $\sigma \in S_n$ denotes the action of the permutation group on the arguments of the field:
\be\label{rev03}
\phi^\sigma(g_a) := \phi(g_{\sigma(a)}).
\ee
Moreover, $[dg_a]$ is shorthand for the normalized measure on $\cG^{\otimes n}$, while $\cK$ and $\cV$ are the kinetic and potential operators, respectively. We further require the invariance of the field under the gauge group:
\be\label{rev04}
\phi(g_a)\rightarrow \phi(g_ag) = \phi(g_a),\qquad\forall g\in \cG\,.
\ee
One can realize this symmetry explicitly in a neat fashion by a simple group averaging:
\be\label{rev05}
\phi(g_a) := \int_{\cG}dg \;{\varphi}(g_ag),
\ee
where $\varphi$ is an auxiliary non-invariant field.

We define the partition function based on this action in the straightforward way\footnotemark:
\be\label{rev06}
\cZ = \int \cD\phi \; e^{-\cS_\lambda[\phi]} =  \sum_N \sum_{\Delta_N} \frac{\lambda^N}{sym[\Delta_N]} \cZ[\Delta_N].
\ee
\footnotetext{
Here we have taken $e^{-S}$ in the path integral as in statistical physics, but we can also define the partition function with $e^{iS}$ which would truly quantize the group field theory and which would be more natural from a ``third quantization" point of view. Using one or the other depends on what the purpose of the partition function. Although this is an important question for the interpretation of group field theory in general, this issue is not relevant to the discussion in the present paper.
}
where in the second equality we have performed a perturbative expansion in $\lambda$. In that case, $N$ is the order in $\lambda$, $\Delta_N$ denotes the Feynman diagrams with $N$ vertices, while $sym[\Delta_N]$ is the symmetry factor associated to $\Delta_N$.

The Feynman diagrams are identified as 2-complexes defining space-time (pseudo-)triangulations (or more generally cellular decompositions).
As in matrix models, the interaction term generates the fundamental building blocks of the discrete manifold, while the propagator glues them together along their boundary. Then for a given Feynman diagram $\Delta$, its evaluation $\cZ[\Delta]$ defines the spinfoam amplitude associated to the corresponding triangulation.


\medskip

It is illuminating to illustrate these general concepts in terms of a specific model: the GFT for topological $BF$ theory in $n$-dimensions. This is highly relevant to the spinfoam program because topological BF theory is the starting point of the whole construction of spinfoam models. Indeed the spinfoam models for BF theory are the only ones which have been shown to provide a consistent and correct quantization of the theory. Moreover, 3d gravity is exactly a topological BF theory while 4d general relativity can be formulated as a BF theory with non-trivial potential.
%

Thus, for BF theory, the interaction term generates $n$-simplices, while the propagator glues them together along shared $(n-1)$-simplices. The fundamental operators are:
\be\label{rev07}
\ba{rcl}
\cK(g_a, \tg_a) &=& \dsty \int dg\prod_{a = 1}^{n}\delta(\tg_{\bar{\iota}(a)}^{-1}\,g_a \,g),\\ [0.5cm]
\cV(\{g_{ab}\}) &=& \dsty \int[dg_a] \prod_{a = 1}^{n}\prod_{b\,: \,b> a} \delta(g_{ba}^{-1}\, g_b\, g_a^{-1}\, g_{ab})
\ea
\ee
These choices for $\cK$ and $\cV$ are usually referred as ``trivial" in the GFT framework. For instance, $\cK$ is simply the projector on gauge-invariant fields, that is the identity on the space of gauge-invariant fields \eqref{rev04}.
%
This choice is appropriate for topological models (of BF type) and it is a widely used one also in non-topological models based on constraining/deforming topological ones. The interaction term $\cV$ here simply identifies the group elements $g_{ab}$ and $g_{ba}$ up to gauge transformations. In this sense, we can call it trivial. The standard prescription for GFTs is to keep a trivial propagator $\cK$ while encoding all the non-trivial information and dynamics in the interaction vertex $\cV$.

With these definitions, the propagator is $\cP := \cK^{-1} = \cK$ and the amplitude for a specific Feynman graph is:
\be\label{re08}
\cZ[\Delta]  =  \int [dg] \prod_e \cP \prod_v \cV = \int \prod_e dg_e \prod_{l} \delta(G_l) .
\ee
In the final expression, $l$ denotes the loops in the Feynman graph, which are dual in the topological sense to the $(n-2)$-simplices of the discrete manifold. Moreover,  $G_l \equiv \prod_{e\in l} g_{e}^{\epsilon(e, l)}$, where $\epsilon(e,l) = \pm 1$ depends on the relative orientation of $e$ and $l$. In the end, we recognize the discretized quantum amplitude for BF theory, in its lattice gauge theory formulation, as expected.

So, in this BF case, the free theory has trivial dynamics and it does not contain any interesting information on the behavior of the full theory.  In particular, the only classical solution is $\phi = 0$, i.e. the ``no-space-time" configuration. As noted earlier, the entire non-trivial dynamics of the theory, imposing the (Hamiltonian) flatness constraint (as a graph-changing operator) on geometry and at the same time governing topology changing processes, lies in the GFT interation term.
This is why we will study the variations of the GFT around non-trivial classical solutions of the full GFT. We will see that the effective kinetic term defining the new free theory actually carries non-trivial information on the GFT dynamics; moreover, it defines a graph-preserving, and thus topology-preserving, quantum dynamics for geometry, leaving graph-chaging and topology-changing processes to be generated by the (effective) interaction term.

\subsection{Spin Network Observables}

For field theories in general, physical observables are deemed to be functions of the fields that are invariant under the (gauge) symmetries of the theory.  In the class of GFT models dealt with above, the fundamental field is a scalar $\phi$, which is invariant under the action of the symmetry (rather then covariantly transforming with respect it).  Thus rather arbitrary functions of the field suffice to encode acceptable physical observables.

Consider an arbitrary product of $V$ fields:
\be\label{net00}
\psi^{\{\phi_v\}}_{\{\sigma_v\}}(g_{vw}) = \prod_v^V \phi_v^{\sigma_v}(g_{vw}),
\ee
This tensor product $\psi^{\{\phi_v\}}=\bigotimes_v  \phi_v$ represents an arbitrary multi-particle state for the group field theory.
There is a neat graphical interpretation of such an entity:  a vertex $v$ describes the field, while the edges $vw$ incident at that vertex denote the arguments of this field.  Thus, one views each $\phi_v^\sigma(g_1, \dots, g_n)$ as an n-valent vertex. The element $\sigma\in S_n$ defines an ordering of the edges around the vertex (as when projected onto a plane).  Moreover, there is no coupling among the fields, thus there is no sense in which these vertices are connected to  each other in any manner. The index $w$ merely denotes the open end of the edge $vw$.

\begin{figure}[h]
\begin{center}
\includegraphics[height=30mm]{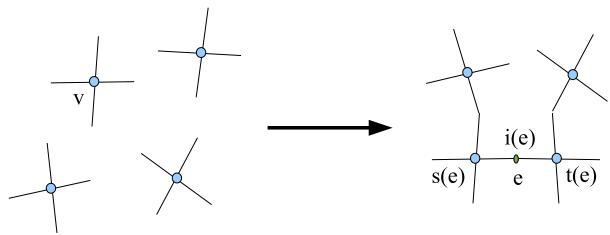}
\caption{The functional $\psi^{\{\phi_v\}}$ consisting in group fields $\varphi_v$ at unrelated vertices $v$ and the functional $\psi_{\Gamma}^{\{\phi_v\}}$ constructed by gluing these vertices along the edges of a graph $\Gamma$. For a given edge $e$, the two group field living at the source vertex $s(e)$ and target vertex $t(e)$ are glued by an intermediate (fiducial) vertex $i(e)$.\label{spinnet}}
\end{center}
\end{figure}

Then, from the quantum gravity viewpoint, a particularly interesting subclass of observables are those that can be labeled by connected graphs $\Gamma$:
\be\label{net01}
\psi_{\Gamma,\{\sigma_v\}}^{\{\phi_v\}}(G_e)=
\int [dg_{vw}]\,
\prod_v \phi_v^{\sigma_v}(g_{vw})\,
\prod_e \delta(g_{t(e)i(e)}^{-1}g_{s(e)i(e)}G_e^{-1}).
\ee
We note that the first product is exactly that occurring in the observable $\psi^{\{\phi_v\}}$ above.  The second product serves to couple these fields, in effect by gluing pairs of edges at their a priori free endpoints.  More precisely, $s(e)$ and $t(e)$ denote the source and target vertices of the edge $e$ in  the graph $\Gamma$.
The index $i(e)$ is the intermediate index of the group elements at the vertices $s(e)$ and $t(e)$ which allows to glue the group elements along the edge $e$. It can be thought as an intermediate vertex along the edge $e$\footnote{One can also define a slightly different but equivalent gluing procedure, giving the same spin network functionals from products of GFT fields. Instead of inserting a delta function per edge constraining the arguments of the GFT fields, one can impose, by projection, an extra \lq gluing symmetry\rq by considering only those products of fields whose arguments referring to the (would be) same edge $e$ of the closed spin network graph are invariant under translations by the same group element $h_{i(e)}$.}. This can be seen on fig.\ref{spinnet}.
%
Ultimately, in the quantum gravity language, the coupling term imposes that the holonomies along the two segments of the edge compose (under group multiplication) to a holonomy $G_e$ for the whole edge.

From the form of the observable, the symmetry of the field $\phi$ under the diagonal action of the group ensures the following symmetry for $\psi_{\Gamma,\phi}$:
\be\label{net02}
\psi_{\Gamma}^{\{\phi_v\}}(G_e)=
\psi_{\Gamma}^{\{\phi_v\}}(h_{s(e)}G_eh_{t(e)}^{-1})\,,
\ee
where we have removed the subscripts $\{\sigma_v\}$ to lighten the notations. Reversely, any function satisfying this gauge invariance can be written as a multi-particle states \eqref{net01} of the group field theory.

This symmetry is very familiar from the spin network observables arising in the spin foam approach.  In that context, one has gauge invariant functions of the connection with support on graphs, that is, functions of the form $\psi_{\Gamma}(G_e)$ with the same symmetry \eqref{net02}.  Here, one has gauge invariant {\it functionals} of the connection with support on graphs.
What is even more appealing is that  in the quantum gravity setting, these spin networks functions form a basis for the kinematical state space.  Indeed, the GFT functionals can be expanded in terms of these spin-network functions.    In other words, the GFT observables may be viewed as functionals of these states, i.e. functionals of the same wave functions defining quantum states of geometry in canonical loop quantum gravity. This is in agreement with the interpretation of GFTs as second quantiztaions of canonical loop quantum grvaity \cite{gft1,gft_dan,gft_3rd}.

\medskip

Following this, we can introduce a natural set of observables for the group field theory:
\be
\psi_{\Gamma}[\phi](G_e)
\,=\,
\int [dg_{vw}]\,
\prod_v \phi^{\sigma_v}(g_{vw})\,
\prod_e \delta(g_{t(e)i(e)}^{-1}g_{s(e)i(e)}G_e^{-1}),
\ee
which is a polynomial function of the group field $\phi$.

These observables have  a natural field theoretic interpretation, namely that:
\be\label{net04}
\ba{rcl}
\la \psi_{\Gamma}[\phi](G_e)\ra &=& \dsty \f1Z\int \cD \phi \;\psi_{\Gamma}[\phi](G_e)\; e^{-S_\lambda[\phi]},\\[0.3cm]
\la \psi_{\Gamma^1}[\phi](G^1_e)\;\psi_{\Gamma^2}[\phi](G^2_e)\ra  &=& \dsty \f1Z\int \cD \phi \;\psi_{\Gamma^1}[\phi](G^1_e)\;\psi_{\Gamma^2}[\phi](G^2_e)\; e^{-S_\lambda[\phi]}, etc.
\ea
\ee
define the probability amplitude for the boundary state $\psi_{\Gamma}(G_e)$ and transition amplitude between two spin network states $\psi_{\Gamma^1}(G^1_e)$ and $\psi_{\Gamma^2}(G^2_e)$, respectively. Indeed, if we expand these correlations perturbatively in the coupling $\lambda$, we recover the standard sum over all spinfoam structures compatible with the boundary graph(s) (see e.g. \cite{gft0,gft_alej} for more details). Here, we parameterize the boundary data with a graph $\Gamma$ and group elements $G_e$ (up to gauge transformations) on the graph edges. If we want to go to the standard spin network basis, we just have to do harmonic analysis on the group $\cG$ and go to boundary data labeled by representations and intertwiner states\footnote{One can also use a spinorial representation of the same functions \cite{spinor}, or go to a triad (flux) representation \cite{flux}, using the non-commutative group Fourier transform \cite{fourier,fourier2}}.

\medskip

Finally,  we notice that all the terms in the GFT action are given as spin network observables of the above type.  For the GFT formulation of BF theory, the kinetic term corresponds to the $\Theta$-graph (made of two vertices), while the potential corresponds to an $n$-vertex graph (corresponding to a $n$-simplex). More precisely, the kinetic and potential terms are given by the evaluation of the corresponding spin network functionals at the identity $G_e=\id$.
Under renormalization, we may expect effective terms of the type $\psi_{\Gamma}[\phi](G_e)$ for other graphs $\Gamma$ to enter the effective group field theory action as quantum corrections (or counter-terms). Of course, we expect terms given by the evaluation at the identity $G_e=\id$ as before, but effective terms with evaluations on more general group elements or derivative terms would be probably a sign of non-trivial curvature corrections.

\subsection{Non-Trivial Backgrounds and Effective Action}

As mentioned earlier, the standard GFT action is usually prescribed with a trivial kinetic operator and hence propagator. It does not contain any derivative terms and is a simple mass term. The induced equation of motion is
\be\label{triv01}
\frac{\delta \cS_\lambda[\phi]}{\delta \phi(g_a)}  = \phi(g_a) - \lambda \int [dg_{ab}] \frac{\delta}{\delta \phi(g_a)}\Big(\prod_{a=1}^m \phi^{\sigma_a}(g_{ab})\Big)\,\cV(\{g_{ab}\}) = 0.
\ee
The trivial classical solution is obviously  $\phi=0$.
This is to be compared to the equation of motion of the {\it free} theory defined by solely considering the kinetic term (and discarding the interaction term):
\be\label{triv03}
\frac{\delta \cS^{free}[\phi]}{\delta \phi(g_a)}  = \phi(g_a) = 0,
\ee
whose {\it only} classical solution is $\phi=0$.

To go further, it is rather convenient to rescale the field so that $\lambda$ disappears from the equations of motion:  $\phi \rightarrow \lambda^{-\frac{1}{m - 2}}\phi$.  The action under this redefinition transforms as:
\be\label{triv04}
\cS_\lambda[\phi] \rightarrow \cS_{\lambda}[\lambda^{-\frac{1}{m - 2}}\phi] = \lambda^{-\frac{2}{m - 2}}\cS[\phi]\,.
\ee
To maintain a certain level of generality, let us assume that we have some non-trivial solution:  $\phi = \phi_0$ such that $\dsty\frac{\delta \cS}{\delta \phi}\Big|_{\phi_0} = 0$. Then one may rewrite any field configuration as $\phi = \phi_0 + \varphi$.   Now we can substitute this decomposition into the action to obtain (schematically):
\be\label{triv05}
\lambda^{-\frac{2}{m - 2}} \cS[\phi] = \lambda^{-\frac{2}{m - 2}}\left[\cS[\phi_0] + \frac{\delta\cS}{\delta \phi}\Big|_{\phi_0}\;\varphi + \frac{1}{2}\frac{\delta^2\cS}{\delta \phi^2}\Big|_{\phi_0}\varphi^2 + \sum_{a = 3}^{m}\frac{1}{m!}\frac{\delta^a\cS}{\delta \phi^a}\Big|_{\phi_0}\varphi^a\right].
\ee
Naturally, the $\cS[\phi_0]$ term may be dropped since it is constant and does not affect the classical dynamics. What is more, in the quantum theory, it cancels in the evaluation of normalized correlation functions.  The second term is zero since $\phi_0$ is a classical solution.  So, it is the third term and onwards that contain the effective dynamics of the field $\varphi$:
\be\label{triv06}
\cS_{\phi_0}[\varphi] := \lambda^{-\frac{2}{m - 2}}\left[\f12\frac{\delta^2\cS}{\delta \phi^2}\Big|_{\phi_0}\varphi^2 + \sum_{a = 3}^{m}\f1{m!}\frac{\delta^a\cS}{\delta \phi^a}\Big|_{\phi_0}\varphi^a\right].
\ee
This is just a simple recasting of the theory in terms of different variables, i.e. we are not changing the non-perturbative dynamics of the model. However, and this is the power and motivation of the approach, it amounts to considering, in perturbation theory, the dynamics around a new, non-trivial phase of the theory, and one that can be reached only non-perturbatively, from the point of view of the \lq no- space\rq vacuum $\phi=0$. This simply follows from the fact that the field  $\varphi$ is a perturbation around the classical background solution $\phi_0$.
%
%

Let us start with analyzing at the free theory. Looking at just the quadratic term, one notes immediately that it now contains a non-trivial kinetic operator. Thus, now even in the free theory, one has non-trivial propagation.  Thus, in analogy with other field theories it is tempting to consider it as the Hamiltonian operator for the corresponding single particle theory or, in the context of gravity, as the Hamiltonian constraint operator:
\be\label{free01}
\cS_{\phi_0}^{free}[\varphi] = \int \varphi \;\hat{\cH}\; \varphi \quad\quad \textrm{where}\quad\quad \hat{\cH} = \frac{\delta^2\cS}{\delta\phi^2}\Big|_{\phi_0} .
\ee
This operator $\hat{\cH}$ contains the trivial contribution coming from the original kinetic term but more importantly it also contains a non-trivial contribution coming from the original interaction term of the GFT action. Thus it carries non-trivial information about the {\it full} dynamics of the theory. The equation of motion of this free theory is now simply $\hat{\cH}\; \varphi=0$ and, thus, at this level, we are interested in the zero modes in the spectrum of $\hat{\cH}$. 

One can also go further and let $\hat{\cH}$ act on the multi-particle states of the theory, that is on the spin network states:
\be\label{free02}
\hat{\cH} \vartriangleright \psi_{\Gamma,\{\sigma_v\}}^{\{\phi_v\}}(G_e)
\,=\,
\int [dg_{vw}]\,  \sum_a \left[\hat{\cH}\vartriangleright\varphi_a^{\sigma_a}(g_{aw})
\prod_{v:v\neq a} \varphi_v^{\sigma_v}(g_{vw})\right]
\prod_e \delta(g_{t(e)i(e)}^{-1}g_{s(e)i(e)}G_e^{-1}).
\ee
Still looking for the zero modes of this operator, this means that we have defined a constraint operator acting on spin network states. This is our proposal to define a tentative Hamiltonian constraint for LQG's spin networks from GFT. In order to assert the physical relevance of our procedure, we would have to test it on some specific spinfoam model(s). This is what we'll do in the next section, where we will apply it to the GFT for 3d quantum gravity.

Next, we need to investigate the full theory, beyond its kinetic term.
From this more general persepctive,  $\hat{\cH}^{-1}$ defines the propagator of the GFT, which enters the evaluation of the Feynman diagrams for the perturbative expansion of the path integral of the theory.  Thus, we are interested for the interacting theory in the complete spectrum of $\hat{\cH}$, as we expect states to be allowed to go \lq off-shell\rq with respect to the constraint.  Moreover, as we know from standard quantum field theory, it is essential to know the full spectrum of the propagator in order to study the properties of the field theory and its renormalization.

%
%
%

\section{The 3d GFT formulation of topological BF theory}

\subsection{The Boulatov model and flat Solutions}

Now, it is time to specialize to an explicit example to see how our strategy plays out. We shall apply it to Boulatov's group field theory for quantum BF theory in 3 dimensions with gauge group $\SU(2)$ \cite{boulatov} (equivalently, 3d Riemannian quantum gravity), whose Feynman amplitudes give the spinfoam amplitudes of the Ponzano-Regge model.

We choose a compact semi-simple Lie group $\cG$ and consider invariant fields on $\cG^{\times 3}$:
$$
\phi(g_1,g_2,g_3)=\phi(g_1g,g_2g,g_3g),\qquad\forall g\in\cG.
$$
Explicitly, the action for Boulatov's GFT is:
\be\label{bou01}
\cS_\lambda[\phi] = \f12\int [dg] \; \phi(g_1,g_2,g_3)\,\phi(g_3,g_2,g_1) - \f{\lambda}{4!}\int [dg] \; \phi(g_1, g_2,g_3)\, \phi(g_3,g_5,g_4)\,\phi(g_4,g_2,g_6)\,\phi(g_6,g_5,g_1).
\ee
The equations of motion $\dfrac{\delta \cS}{\delta \phi} = 0$ take the form:
\be\label{bou02}
\phi(g_3,g_2,g_1) - \frac{\lambda}{3!} \int [dg] \phi(g_3, g_5,g_4)\,\phi(g_4,g_2,g_6)\,\phi(g_6,g_5,g_1) = 0
\ee
To this equation, there exists a family of classical solutions labeled by functions $f: \cG \rightarrow \R$
identified in \cite{gft_winston}, namely:
\be\label{bou03}
\phi_f(g_1,g_2,g_3) = \sqrt{\f{3!}\lambda} \int dg \;\delta(g_1 h)\, f(g_2 h)\, \delta(g_3 h), \quad\quad\textrm{provided}\quad\quad \int dg \; f(g)^2 = 1.
\ee
There exists of course other classical solutions (see e.g. \cite{gft_flo}) as well as approximate solutions \cite{danielelorenzo}, but we will focus on the family of solutions defined above and referred to as ``flat solutions".

With these solutions at our disposal, we may perturb around them as specified earlier using $\phi = \phi_f + \varphi$.  The effective action for the field $\varphi$ is:
\bes\label{bou04}
\cS_{\phi_f}[\varphi]
&=&
\f12\int [dg]^3\, \varphi(g_1,g_2,g_3)\,\varphi(g_3,g_2,g_1)\nn\\
&&
-\int[dg]^4\,f(g_2g_1^{-1})f(g_5g_1^{-1})\varphi(g_1,g_2,g_6)\varphi(g_6,g_5,g_1)
-\f12\int[dg]^4\,f(g_2g_1^{-1})\varphi(g_1,g_5,g_4)f(g_2g_4^{-1})\varphi(g_6,g_5,g_1)
\nn\\
&&- \f {\sqrt{\lambda}}{\sqrt{3!}} \int [dg]^5 \, f(g_2 g_3^{-1})\, \varphi(g_3,g_5,g_4)\,\varphi(g_4,g_2,g_6)\,\varphi(g_6,g_5,g_3) \nn\\
&&- \f \lambda{4!} \int [dg]^6\,  \varphi(g_1,g_2, g_3)\, \varphi(g_3,g_5,g_4)\,\varphi(g_4,g_2,g_6)\,\varphi(g_6,g_5,g_1),
\ees
where the corrections to the kinetic term and the new cubic interaction comes from the original interaction vertex of the GFT. Focusing on the kinetic term, we see that the free theory does not depend on the coupling $\lambda$ at all and is given by the quadratic action:
\be
\label{kin}
\cS_{\phi_f}^{free}[\varphi] = \f12\int [dg]\; \varphi(g_1,g_2,g_3)\, {\cH}(g_1,g_2,g_3;\tg_1,\tg_2,\tg_3) \,\varphi(\tg_1,\tg_2,\tg_3)
\ee
with the kinetic operator is
\be\label{bou05}
{\cH}(g_a;\tg_b) = \delta(g_3\tg_1^{-1})
\left[
\delta(g_2\tg_2^{-1})
-\delta(g_2\tg_2^{-1})\,\int dh \, f(hg_1^{-1})\, f(hg_3^{-1})
- 2 \,f(g_2g_1^{-1})\,f(\tg_2g_1^{-1})
\right]
\delta(g_1\tg_3^{-1}).
\ee
%

\subsection{The Spectrum of the Propagator}


We define the operator $\hcH$ acting on invariant fields following the formula for the free theory given above:
\bes
\hcH\,\varphi(g_1,g_2,g_3)
&=&
\varphi(g_1,g_2,g_3)\left(
1-\int dh\, f(hg_1^{-1})f(hg_3^{-1})
\right)\\
&&
-2f(g_2g_3^{-1})\,\int d\tg_2\, \varphi(g_1,\tg_2,g_3)f(\tg_2 g_3^{-1})\,. \nn
\ees
With the standard scalar product, $\la\varphi|\widetilde{\varphi}\ra=\int [dg]^3 \overline{\varphi}(g_a)\,\widetilde{\varphi}(g_a)$, it is straightforward to check that this operator is Hermitian (since $f$ is a real function). Moreover, for a field satisfying the reality condition $\overline{\varphi}(g_1,g_2,g_3)=\varphi(g_3,g_2,g_1)$, the kinetic term \eqref{kin} of the free theory is exactly given by the scalar product $\la\varphi|\hcH|\varphi\ra$. Thus the equation of motion of our free theory defined by the quadratic term of the effective action $S_{\phi_f}$ is simply $\hcH\,\varphi=0$.
We will now look for the eigenstates satisfying $\hcH\,\varphi\,=\mu\varphi$ and fully diagonalize the operator $\hcH$.

%

We consider the action of the operator $\hcH$ on the space on invariant fields, $\varphi(g_1,g_2,g_3)=\varphi(g_1g,g_2g,g_3g)$ for all $g\in\cG$. Such invariant functions can be looked at as functions of the gauge invariant combinations $g_1g_3^{-1}$ and $g_2g_3^{-1}$. Thus we acting with $\hcH$ on the Hilbert space $L^2(\cG^{\times 3}/\cG)\sim L^2(\cG^2)$. A basis of functions on $L^2(\cG^2)$ is given by tensor product states such $\varphi(g_1,g_2,g_3)=\rho(g_1g_3^{-1})A(g_2g_3^{-1})$ or in short $\varphi=\rho\otimes A$ . It turns out that such simple states already diagonalize $\hcH$. 
We distinguish two cases:
\begin{itemize}

\item $\mathbf{A\bot f}$: if $\la f|A\ra=\int dh\,f(h)A(h)=0$, the action of $\hcH$ on $\varphi(g_1,g_2,g_3)=\rho(g_1g_3^{-1})A(g_2g_3^{-1})$ simplifies to
    \be
    \hcH\,\rho(g_1g_3^{-1})A(g_2g_3^{-1})
    \,=\,
    \rho(g_1g_3^{-1})A(g_2g_3^{-1})\,\left(
    1-\int dh\, f(h)f(hg_1g_3^{-1})
    \right).
    \ee
    From this, it clear that taking $\rho(g)=\delta_G(g)=\delta(gG^{-1})$ for a fixed group element $G\in\cG$ will diagonalize this action. Then the tensor product states $\varphi=\delta_G\otimes A$ with $A\bot f$ are eigenstates of $\hcH$:
    \be
    \hcH\,\delta(g_1g_3^{-1}G^{-1})A(g_2g_3^{-1})
    \,=\,
    \left(    1-\int dh\, f(h)f(hG)\right)\,
    \delta(g_1g_3^{-1}G^{-1})A(g_2g_3^{-1}).
    \ee
    The corresponding eigenvalues are $\mu=( 1-\int dh\, f(h)f(hG))$ and do not depend on the choice of the function $A$. Since $\int f^2=1$ is normalized, the Cauchy-Schwarz inequality ensures that $\int dh\, f(h)f(hG)$ is bounded by 1 in absolute value, thus we have $\mu\in[0,+2]$\,.

    The lowest eigenvalue $\mu=0$ is reached when saturating the Cauchy-Schwarz inequality, that is for $G=\id$. In that case, the eigenvector $\varphi(g_1,g_2,g_3)=\delta(g_1g_3^{-1})A(g_2g_3^{-1})$ is just the flat classical solution $\phi_A$ (up to a proportionality factor).

\item $\mathbf{A\propto f}$:  In the case that $A=f$ (since the operator is linear, the proportionality factor is irrelevant here), the action also simplifies:
    \be
    \hcH\,\rho(g_1g_3^{-1})f(g_2g_3^{-1})
    \,=\,
    \rho(g_1g_3^{-1})f(g_2g_3^{-1})\,\left(
    1-\int dh\, f(h)f(hg_1g_3^{-1})-2\int f^2)
    \right).
    \ee
    Using the normalization $\int f^2=1$ and once again taking the ansatz $\rho(g)=\delta_G(g)$, we obtain the remaining eigenstates:
    \be
    \hcH\,\delta(g_1g_3^{-1}G^{-1})f(g_2g_3^{-1})
    \,=\,
    -\left(1+
    \int dh\, f(h)f(hG)
    \right)\,
    \delta(g_1g_3^{-1}G^{-1})f(g_2g_3^{-1}),
    \ee
    which gives eigenvalues $\mu=\,-1- \int dh\, f(h)f(hG)\,\in[-2,0]$.

    All the eigenvalues in this sector are lower than in the previous section $A\bot f$. And once again, the lowest eigenvalue is reached for $G=\id$ and $\rho=\delta$, whose corresponding eigenstate of our Hamiltonian constraint $\hcH$ is the state $\varphi(g_1,g_2,g_3)=\delta(g_1g_3^{-1})f(g_2g_3^{-1})$, which is (up to an irrelevant factor) the initial flat solution $\phi_f$ around which we have expanded the GFT.

\end{itemize}

To summarize: we have checked that the spectrum of our (Hermitian) constraint operator $\cH$ is bounded both from above and from below; the eigenvalues are $\pm 1-\int dh \, f(h)f(hG)$ and are thus parameterized, as the corresponding eigenfunctions, by an arbitrary group element $G$; the spectrum is therefore generically continuous and it depends explicitly on the background classical solution $\phi_f$ defined by the function $f$.

\medskip

It is also interesting to compare the constraint operator we have obtained, and its spectrum, to the standard $p^2-m^2$ of a scalar field theory. Indeed the quantity $-\int dh \, f(h)f(hG)$ can be roughly identified as the momentum squared $p^2$. This was already noticed in \cite{gft_winston} where the kinetic operator of the effective non-commutative field theory for a scalar field coupled to 3d quantum gravity (as derived in \cite{fourier}) can be in such fashion.

More precisely, let us take the gauge group $\cG=\SU(2)$ and assume that $f$ is a central function, $f(hgh^{-1})=f(g)$. Then $f$ can be expanded over the characters $\chi_j$ of the irreducible representations of $\SU(2)$ labeled by the spin $j\in\N/2$:
\be
f(g)=\sum_j f_j \chi_j(g),
\qquad
\sum_j f_j^2 =1,
\qquad
\int dh \, f(h)f(hG)=\sum_j f_j^2 \f{\chi_j(G)}{d_j},
\ee
where the coefficients $f_j$ are real and the factors $d_j=\chi_j(\id)=(2j+1)$ is the dimension of the $\SU(2)$-representation of spin $j$.
%
%
A simple manipulation allows to write:
$$
-\int dh \, f(h)f(hG)=-1+\sum_{j} f_j^2 \left(1-\f{\chi_j(G)}{d_j}\right),
$$
due to the normalization condition $\sum_j f_j^2 =1$.
Now, since the characters reach their absolute maximum in the identity, $\chi_j(\id)=d_j$ by definition, then the series in the equation above is always positive and vanishes at the identity $G=\id$.
Thus we can write:
\be
-\int dh \, f(h)f(hG)=P^2(G)-1
\qquad\textrm{with}\quad
P^2(G)\,\equiv\,\sum_j f_j^2 \left(1-\f{\chi_j(G)}{d_j}\right),
\ee
where the shift $-1$ can be interpreted as a mass term. This kinetic term can truly be written through a (group) Fourier transform as a Laplacian operator in term of an actual 3-momentum on the non-commutative $\R^3$ dual to the group manifold $\cG=\SU(2)$ \cite{fourier,fourier2}.

\medskip

Thus, through our procedure we have recovered a non-trivial kinetic term which can be interpreted and used in analogy with the usual $(p^2-m^2)$ of standard quantum field theory. This is interesting also recalling that the Hamiltonian constraint of geometrodynamics is indeed of the form of a (functional) Klein-Gordon-type quadratic operator on superspace. However, the analogy must be taken with care, because the situation is different in three spacetime dimensions, which is actually the case we are dealing with here. This point being understood, the main interest in the above result is that it confirms that the kinetic operator we have defined is non-trivial but manageable, and that the corresponding propagator has a non-trivial spectrum and it is thus directly amenable to more standard constructive techniques in the context of GFT renormalization \cite{gft_renorm}.

\subsection{The Effective Hamiltonian Constraint}

We are ready to realize the action of this effective Hamiltonian constraint on a generic spin network state for graph $\Gamma$, given, as we have seen, by the tensor product of group field $\varphi_v$ associated to each of the vertices $v$ of the graph:
\be
\psi_{\Gamma}(G_e) = \int [dg_{vw}]\,
\prod_{v} \varphi^{\sigma_v}(g_{vw})\,
\prod_e \delta(g_{t(e)i(e)}^{-1}g_{s(e)i(e)}G_e^{-1}),
\ee
so that a spin network state is interpreted as a multi-particle state (in the Fock space) of the GFT.
Then we let the linear operator $\hcH$ act as expected on the tensor product $\psi_\Gamma=\bigotimes_v \varphi_v$ as earlier in \eqref{free02}:
\be
\hat{\cH} \vartriangleright \psi_{\Gamma}(G_e) = \int [dg_{vw}]\,  \sum_a \left[\hat{\cH}\vartriangleright\varphi^{\sigma_a}(g_{aw})
\prod_{v:v\neq a} \varphi^{\sigma_v}(g_{vw})\right]
\prod_e \delta(g_{t(e)i(e)}^{-1}g_{s(e)i(e)}G_e^{-1}),
\ee
$\psi_\Gamma$ is therefore an eigenstate  of $\hcH$ if (and only if) the fields $\varphi_v$ are all eigenvectors of the Hamiltonian operator.

For any choices of $\varphi_v$, it is straightforward compute the associated spin network functional by explicitly performing the integrals over the gauge group $\cG$. Here we focus on identifying and interpretating the state corresponding to the lowest eigenvalue of the effective Hamitonian constraint operator, which we call for simplicity \lq\lq ground state\rq\rq.

For instance, the ground state on a graph $\Gamma$ will be given by taking the ground state of the group field everywhere, i.e $\varphi_v=\phi_f$ for all vertices $v$ as we have derived in the previous section. A minor point is that the eigenvalue of $\hcH$ associated to $\varphi=\phi_f$ is $-2$, so that it might be a better convention to shift $\hcH$ by $2$ in order to define the ground state as having vanishing eigenvalue (then, solutions of the Hamiltonian constraint equation would have eigenvalue $2$).

\begin{figure}[h]
\begin{center}
\includegraphics[height=20mm]{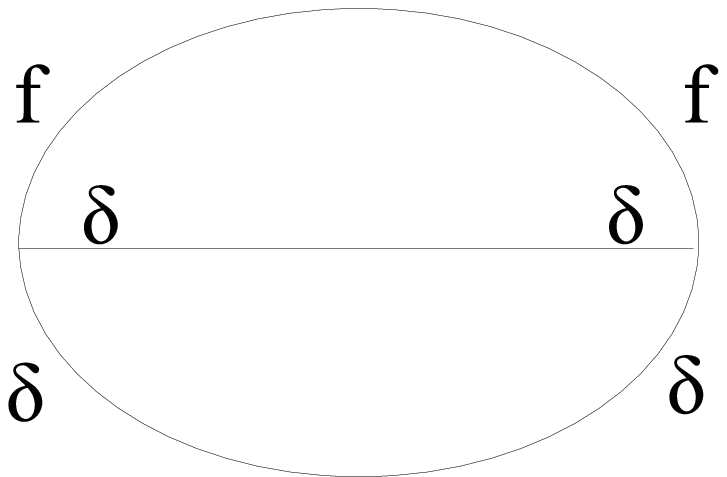}
\hspace*{10mm}
\includegraphics[height=20mm]{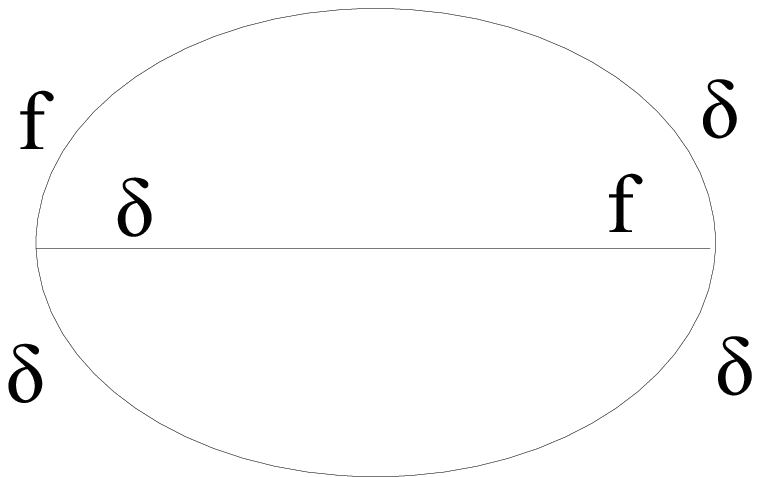}
\caption{The $\Theta$-graph with two vertices and three edges linking them: the two possibilities for defining the spin network functional made from $\phi_f$ group field insertions at the two vertices but with different choices of permutations at each vertex.\label{Thetagraph}}
\end{center}
\end{figure}

Computing the integrals of the product $\prod_v\varphi_v$, we will get a certain combination of $\delta$ functions and convolutions of $f$ of every loop of the graph $\Gamma$. Let us compute this explicitly for the $\Theta$-graph, made of two vertices and three edges linking them. The group field $\varphi=\phi_f=\int \delta f\delta$  has three legs to which are associated twice the $\delta$-distribution and once the function $f$. Thus, we have two possible configurations on the $\Theta$-graph depending on the choice of permutations $\sigma_v$: either the two $f$-insertions are on the same edge (let's say 1) or they are on different edges (let's say 1 and 2). These two possibilities are illustrated on fig.\ref{Thetagraph}.
The corresponding spin network states are easy to compute:
\bes
\psi_\Theta^{(1)}(G_a)
&=&
\delta(G_2G_3^{-1})\,\int dh\,f(h)f(hG_1G_2^{-1})
=\delta(G_2G_3^{-1})\,f\circ f(G_1G_2^{-1}),
 \nn\\
\psi_\Theta^{(2)}(G_a)
&=&
f(G_1G_3^{-1})\,f(G_2G_3^{-1}),
\nn
\ees
where we have assumed that $f$ is central\footnotemark{} for simplicity's sake.
\footnotetext{
For the gauge group $\cG=\SU(2)$, if the function $f$ is invariant under conjugation, $f(g)=f(hgh^{-1})$ then it is automatically invariant under inversion, $f(g)=f(g^{-1})$. Then $\circ$ is simply the conventional convolution product between functions over $\SU(2)$.
}

As expected, these are gauge-invariant functionals invariant under the action of $\cG$ at each vertex of the graph. They assign a certain convoluted power $\delta$ or $f$ or $f\circ f$ or more generally $f^{\circ n}$ to each loop depending on the number of $f$-insertions along that loop.
When we have the $\delta$-distribution,  we are imposing that the holonomy along that loop is trivial and thus that the connection is flat. When we have a non-trivial power $f^{\circ n}$ on a loop, it can be interpreted on the other hand as a topological defect or a non-trivial cycle of the space topology.

\begin{figure}[h]
\begin{center}
\includegraphics[height=30mm]{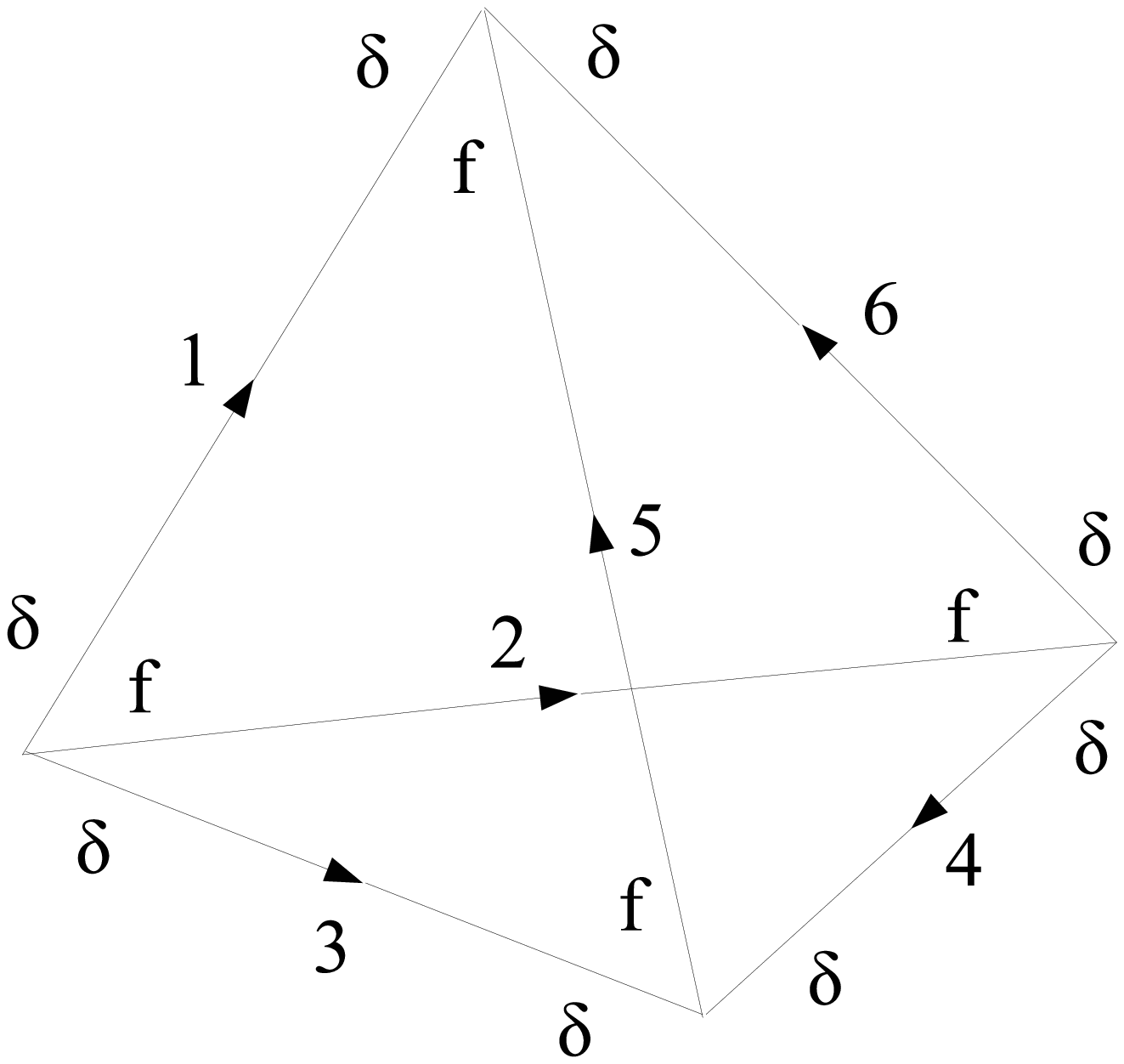}
\caption{The oriented tetrahedron graph with its four vertices and six edges and a particular choice of permutations at the four vertices in order to define the spin network functional.\label{tetra}}
\end{center}
\end{figure}

\begin{figure}[h]
\begin{center}
\includegraphics[height=30mm]{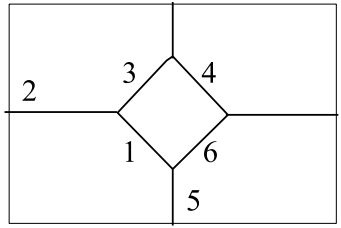}
\caption{The tetrahedron graph (faithfully) embedded on the 2-torus with the edges  2 and 5 wrapped around the two cycles.\label{torus}}
\end{center}
\end{figure}

To assert this interpretation, let us have a try at the tetrahedron graph and consider the choice of permutations as depicted in fig.\ref{tetra}. The corresponding spin network state is straightforward to compute:
\be
\psi_T(G_a)
\,=\,
\delta(G_4G_3^{-1}G_1G_6^{-1})\,f\circ f(G_4G_3^{-1}G_2)\,f\circ f(G_6G_5G_4^{-1}),
\ee
which is interpreted as a flat state on the 2-torus (fig.\ref{torus}) or the  2-sphere with two punctures (topological defects).

Finally, we can also play on the choice of the classical solution $\phi_f$. Indeed, the ground state and more generally all the eigenstates depend on the choice of the function $f$. For instance, if we take the (ill-defined) limit $f\rightarrow\delta/\sqrt{\delta(\id)}$, then the ground state becomes  the completely flat state imposing that the holonomies are trivial along all the loops of the graph. This correctly corresponds to the physical state of topological BF theory for a trivial topology of space. On the other hand, as soon as $f$ is different from the $\delta$-distribution, holonomies become non-trivial and are interpreted as a non-trivial space topology (or topological defects). This describes the ground state of our Hamiltonian constraint. Then we can see that excited states will introduce more and more topological defects and curvature.


\section*{Conclusion \& Outlook}

%
%
%
%

Within the framework of spinfoam models for quantum gravity, we proposed to focus on the free theory defined by the quadratic term of the group field theory and to interpret the kinetic operator as a Hamiltonian constraint. This Hamiltonian constraint $\hcH$ defines the classical equation of motion of the free theory and acts on the GFT field $\varphi$, which represents a single intertwiner of a spin network state for loop quantum gravity. Nevertheless, we have also shown to interpret general spin network states as multi-particle states of the group field theory, as tensor products of the initial field $\varphi$. Then we have a natural action of the Hamiltonian operator $\hcH$ on spin networks.

A standard prescription for group field theories is to start with a trivial propagator (but see the last reference of \cite{gft_renorm}) and to encode all the dynamics (of both geometry, thus the Hamiltonian constraint, and topology) in the interaction term. Nevertheless, even in this case, we follow the proposal from \cite{gft_winston} to expand the group field theory around a non-trivial solution of its classical equations of motion. This background classical solution contains dynamical information from the full theory, since it depends on the original interaction term. Then we have shown that the effective group field theory describing the field variations around that background acquires a non-trivial kinetic term, which we can then interpret as defining an effective Hamiltonian constraint for loop quantum gravity.

We've applied this program explicitly to Boulatov's group field theory for 3d quantum gravity \cite{boulatov}. We have expanded it around the flat classical solutions introduced in \cite{gft_winston} and analyzed in details the spectrum of the induced Hamiltonian constraint operator. We have seen that it can be interpreted as a kinetic operator of the $(p^2-m ^2)$ type. This not only supports the physical relevance of our procedure but allows supports the idea that group field theory can be undertood (at least in certain phases, or for certain perturbation fields) as the momentum representation of a field theory on an actual space-time manifold, which would be obtained through a Fourier transform \cite{gft_flo,gft_aristide,matrix}.

More work is certainly needed. In particular one should investigate different choices of background GFT configurations around which to expand, since this determines a big deal of the effective Hamitlonian constraint. 

Having a non-trivial propagator of the GFT of the $(p^2-m ^2)$ type, thus with a more standard scale dependance, opens the door to an easier use of standard QFT tools to study GFT renormalization, in particular it would be interesting to see how our procedure can be used within the tentative framework that has been recently developed \cite{gft_renorm}. We should also apply our program to spinfoams for 4d quantum gravity and see if we can extract some interesting and physically relevant effective Hamiltonian constraint from the EPRL-FK spinfoam models for instance \cite{fk,ls,eprl,gft_eprl}.

Finally, we believe it would also be interesting to investigate the group Fourier transform \cite{fourier,fourier2} of all our procedure. This can be given two possible interpretations: it may mean going from the momentum representation given by our GFT on a group manifold to a (non-commutative) field theory on a space-time manifold, if the non-commutative dual variables are interpreted directly as coordinates on it \cite{gft_flo,matrix}, or it could mean simply re-writing the same field theory of geometry from connection to triad/flux variables \cite{gft_aristide,diffeos,flux}. In both readings, it would certainly help to understand the physical and geometrical meaning of the induced propagator and effective Hamiltonian constraint.

\section*{Acknowledgments}

EL is partially supported by the ANR ``Programme Blanc" grants LQG-09 and acknowledges financial support from the European Science Foundation (ESF) through the Short Visit travel grants 3578 and 3770. DO gratefully acknowledges financial support from the A. von Humboldt Stiftung, through a Sofja Kovalevskaja Prize.
%

\appendix


\end{document}